\documentclass[a4paper,fleqn]{cas-dc}

\usepackage[numbers,sort&compress]{natbib}
\usepackage{amsmath,amssymb}
\usepackage{graphicx}
\usepackage{booktabs}
\usepackage{url}
\usepackage{subfig}

\graphicspath{{figures/}}
\newcommand{\R}{\mathbb{R}}
\newcommand{\rhoF}{\rho_{\mathrm{FLC}}}

\begin{document}
\let\WriteBookmarks\relax
\shorttitle{ETM-3D: A One-Parameter Multistable Quadratic Flow}
\shortauthors{Khalili Amirabadi and Jalaeian Farimani}
\title[mode=title]{ETM-3D: A One-Parameter Multistable Quadratic Flow with an Explicit Absorbing Set and Intermittent Chaos}

\author[1]{Roya Khalili-Amirabadi}[orcid=0009-0009-5391-5224]
\ead{r-khalili@mail.um.ac.ir}
\affiliation[1]{organization={Department of Applied Mathematics, Ferdowsi University of Mashhad},
                city={Mashhad},
                country={Iran}}

\author[2]{Mohsen Jalaeian-Farimani}[orcid=0000-0002-4910-309X]
\cormark[1]
\ead{mohsen.jalaeian@polimi.it}
\affiliation[2]{organization={Department of Electronics, Information and Bioengineering (DEIB), Politecnico di Milano},
                city={Milan},
                country={Italy}}
\cortext[cor1]{Corresponding author}

\begin{abstract}
We study a one-parameter three-dimensional quadratic flow, denoted ETM-3D, on the benchmark domain $\rho\geq0$. Its homogeneous quadratic triad preserves Euclidean quadratic energy, while the full vector field has constant divergence $-8$. A shifted quadratic Lyapunov function yields an explicit positively invariant absorbing set and therefore global ultimate boundedness. The equilibria reduce generically to a scalar quartic, with a separate singular case at $\rho=\sqrt{5}$. Continuation reveals two saddle-node bifurcations, a generic transcritical bifurcation, a supercritical Hopf bifurcation, and a fold of limit cycles. Stable and unstable period-one branches meet at the fold with their nontrivial Floquet multipliers approaching $+1$. Just beyond this point, a 25-point logarithmic parameter study gives a laminar-length scaling exponent of $-0.551$ (95\% confidence interval $[-0.621,-0.481]$), consistent with type-I intermittency. At larger $\rho$, a chaotic attractor coexists with an asymptotically stable equilibrium and periodic windows occur. Structural invariants are used to distinguish the flow from several classical quadratic chaotic families.
\end{abstract}


\maketitle

\section{Introduction}
Low-dimensional chaotic systems remain useful models for nonlinear analysis because they combine simple governing equations with complex long-term behavior. The Lorenz system is the canonical example, while the Chen and L{\"u} systems, Newton--Leipnik and Rabinovich models, Sprott flows, and the Qi system illustrate the broader diversity of three-dimensional quadratic dynamics \cite{Lorenz1963,ChenUeta1999,Lu2002,LeipnikNewton1981,Rabinovich1978,Sprott1994,Qi2005}. For a new flow to be useful, however, a distinct phase portrait is not enough; boundedness, equilibrium structure, bifurcation mechanisms, and the route to chaos should be reproducible and analytically interpretable. This is particularly relevant for forecasting and identification benchmarks \cite{Gilpin2021} and for structure-preserving modeling based on energy-preserving quadratic operators \cite{Gkimisis2026}.

Chaotic systems are also demanding control problems because sensitivity to initial conditions and uncertainty can degrade robustness and aggravate chattering. Fuzzy sliding-mode approaches have been developed for uncertain chaotic dynamics \cite{Amirabadi2021}, and hybrid fuzzy sliding-mode techniques have subsequently combined chattering reduction with robust performance \cite{Amirabadi2026}. Motivated by the need for a compact but analytically accessible test system, we study
\begin{equation}
\begin{aligned}
\dot{x}&=-2x-\rho y+yz,\\
\dot{y}&=x-y+\rho z-2xz,\\
\dot{z}&=\rho y-5z+xy,
\end{aligned}
\label{eq:system}
\end{equation}
on the benchmark domain
\begin{equation}
\rho\geq0.
\label{eq:domain}
\end{equation}
The vector field is defined for all real $\rho$, and the boundedness theorem remains valid on the full real line; only the bifurcation study is restricted to the nonnegative branch. We call \eqref{eq:system} the \emph{ETM-3D} system. Its quadratic triad $(yz,-2xz,xy)^T$ is not claimed as a new interaction; the contribution concerns the complete one-parameter linear--quadratic architecture and the dynamical organization it produces.

We construct an explicit positively invariant absorbing set, reduce the generic equilibria to a scalar quartic with the singular case $\rho=\sqrt5$ treated separately, and identify two saddle-node bifurcations, a generic transcritical bifurcation, a supercritical Hopf bifurcation, and a fold of stable and unstable limit cycles. A logarithmically spaced laminar-length study tests the type-I intermittency scenario beyond the fold, while larger $\rho$ yields chaos, coexistence with a stable equilibrium, and periodic windows. Structural invariants are also compared with representative classical quadratic families so that the novelty claim rests on the full architecture rather than on the presence of chaos or on the cross terms alone.

\section{Structural Properties and Global Boundedness}
Write \eqref{eq:system} as $\dot{\mathbf{x}}=A(\rho)\mathbf{x}+N(\mathbf{x})$, where $\mathbf{x}=[x,y,z]^T$ and
\begin{equation}
N(\mathbf{x})=[yz,-2xz,xy]^T.
\end{equation}
The quadratic core satisfies
\begin{equation}
\mathbf{x}^{T}N(\mathbf{x})=xyz-2xyz+xyz=0,
\label{eq:energy}
\end{equation}
so it redistributes the Euclidean quadratic energy $E=\frac12\|\mathbf{x}\|^2$ without directly creating or dissipating it. The divergence of the full field is
\begin{equation}
\nabla\!\cdot f=-2-1-5=-8,
\label{eq:div}
\end{equation}
and hence infinitesimal phase volumes contract as $\mathcal{V}(t)=\mathcal{V}(0)e^{-8t}$.

\subsection{A Positively Invariant Absorbing Set}
\textbf{Theorem 1:} For every fixed $\rho\in\R$, all solutions of \eqref{eq:system} are globally ultimately bounded. More precisely, an explicit sublevel set of the function below is positively invariant and globally absorbing.

\emph{Proof:} Define
\begin{equation}
V_\rho=\frac12\left[(x-2\rho)^2+y^2+\bigl(z-(1-\rho)\bigr)^2\right].
\label{eq:V}
\end{equation}
Direct differentiation gives
\begin{align}
\dot V_\rho={}&-2(x-\rho)^2
-\left(y-\frac{\rho(3\rho-1)}{2}\right)^2 \nonumber\\
&-5\left(z-\frac{1-\rho}{2}\right)^2+C(\rho),
\label{eq:Vdot}
\end{align}
where
\begin{equation}
C(\rho)=2\rho^2+\frac{\rho^2(3\rho-1)^2}{4}+\frac{5(1-\rho)^2}{4}.
\label{eq:C}
\end{equation}
Let
\begin{align}
a_\rho&=\left(\rho,\frac{\rho(3\rho-1)}{2},\frac{1-\rho}{2}\right)^T,\\
b_\rho&=(2\rho,0,1-\rho)^T,
\end{align}
and define $d(\rho)=\|a_\rho-b_\rho\|$. Since the quadratic form in \eqref{eq:Vdot} is bounded below by $\|\mathbf{x}-a_\rho\|^2$,
\begin{equation}
\dot V_\rho\leq-\|\mathbf{x}-a_\rho\|^2+C(\rho).
\end{equation}
The reverse triangle inequality and $\|\mathbf{x}-b_\rho\|=\sqrt{2V_\rho}$ imply
\begin{equation}
\dot V_\rho\leq-\left(\sqrt{2V_\rho}-d(\rho)\right)^2+C(\rho).
\label{eq:comparison}
\end{equation}
Therefore, with
\begin{equation}
R(\rho)=\frac12\left[d(\rho)+\sqrt{C(\rho)}\right]^2,
\label{eq:R}
\end{equation}
the compact sublevel set
\begin{equation}
\Omega_\rho=\{\mathbf{x}:V_\rho(\mathbf{x})\leq R(\rho)\}
\label{eq:omega}
\end{equation}
is positively invariant because $\dot V_\rho\leq0$ on its boundary. Moreover, $\dot V_\rho<0$ whenever $V_\rho>R(\rho)$, so every trajectory eventually enters $\Omega_\rho$. \hfill$\square$

The set \eqref{eq:omega} is intentionally conservative; its role is to provide a rigorous global absorbing set, not a sharp estimate of the attractor size.

\section{Equilibria and Local Bifurcations}
The origin $E_0=(0,0,0)$ is an equilibrium for all $\rho$. For a generic nonzero equilibrium define
\begin{equation}
D=2x^2+\rho x+5-\rho^2.
\end{equation}
When $D\neq0$,
\begin{equation}
y=\frac{5x}{D},\qquad z=\frac{x(\rho+x)}{D},
\label{eq:eqyz}
\end{equation}
where $x$ is a real root of
\begin{align}
0={}&8x^4+8\rho x^3+(-6\rho^2+10\rho+35)x^2 \nonumber\\
&+(-4\rho^3+5\rho^2+15\rho)x \nonumber\\
&+2\rho^4-5\rho^3-20\rho^2+25\rho+50.
\label{eq:quartic}
\end{align}
Equation \eqref{eq:eqyz} is not valid in the singular case $D=0$. Direct substitution into \eqref{eq:system} shows that when $\rho^2=5$ an additional equilibrium is
\begin{equation}
E_{\mathrm{sp}}=(0,5,\rho).
\label{eq:specialeq}
\end{equation}
Thus, within the benchmark domain \eqref{eq:domain}, the exceptional case is $\rho=\sqrt{5}$ and $E_{\mathrm{sp}}=(0,5,\sqrt{5})$. If the family is extended outside the scope of this paper, the counterpart $(0,5,-\sqrt{5})$ occurs at $\rho=-\sqrt{5}$.

Numerical continuation on $\rho\geq0$ gives saddle-node events at
\begin{equation}
\rho_{\mathrm{SN1}}\approx2.18875863,\qquad
\rho_{\mathrm{SN2}}\approx3.35516106.
\end{equation}
At the origin,
\begin{equation}
p_0(\lambda)=\lambda^3+8\lambda^2+(17+\rho-\rho^2)\lambda+(10+5\rho-2\rho^2),
\end{equation}
and a simple zero eigenvalue occurs at
\begin{equation}
\rho_0=\frac{5+\sqrt{105}}{4}\approx3.81173769.
\label{eq:rho0}
\end{equation}
Because $f(0,\rho)=0$ for all $\rho$, the saddle-node parameter coefficient vanishes automatically. Let $v$ and $w$ be right and left nullvectors of $D_xf(0,\rho_0)$ normalized by $w^Tv=1$. The two nondegeneracy coefficients for the reduced transcritical normal form are
\begin{align}
a_{\mathrm{TC}}&=w^TQ(v)=\frac{145}{78}+\frac{5\sqrt{105}}{26}\approx3.82954,\\
b_{\mathrm{TC}}&=w^T A_\rho v=\frac{35}{78}+\frac{3\sqrt{105}}{26}\approx1.63106,
\end{align}
where $Q$ denotes the homogeneous quadratic part and $A_\rho=\partial A/\partial\rho$. Both are nonzero, establishing a generic transcritical bifurcation at \eqref{eq:rho0}.

A neighboring nonzero equilibrium undergoes a Hopf bifurcation at
\begin{equation}
\rho_H\approx3.82517466,
\end{equation}
with
\begin{equation}
E_H\approx(-2.17145428,1.27616614,0.42208439)
\end{equation}
and eigenvalues
\begin{equation}
\lambda_1=-8,\qquad \lambda_{2,3}=\pm1.547249826i.
\end{equation}
The crossing speed is $d\operatorname{Re}\lambda_{2,3}/d\rho\approx0.505908261>0$.

For reproducibility, the first Lyapunov coefficient is computed using the Kuznetsov convention \cite{Kuznetsov2004}. In coordinates centered at $E_H$, write
\begin{equation}
\dot u=A_Hu+\frac12B(u,u),
\end{equation}
choose $A_Hq=i\omega_0q$, $A_H^Tp=-i\omega_0p$, and normalize the Hermitian pairing by $\langle p,q\rangle=1$. Since the system is quadratic, the cubic multilinear term vanishes and
\begin{align}
l_1=\frac{1}{2\omega_0}\operatorname{Re}\langle p,&B(\bar q,(2i\omega_0I-A_H)^{-1}B(q,q))\nonumber\\
&-2B(q,A_H^{-1}B(q,\bar q))\rangle.
\label{eq:l1formula}
\end{align}
Evaluation gives
\begin{equation}
l_1\approx-0.03340799<0,
\end{equation}
so the Hopf bifurcation is supercritical and generates a stable periodic orbit for increasing $\rho$.

\begin{figure}[pos=t]
\centering
\includegraphics[width=0.98\columnwidth]{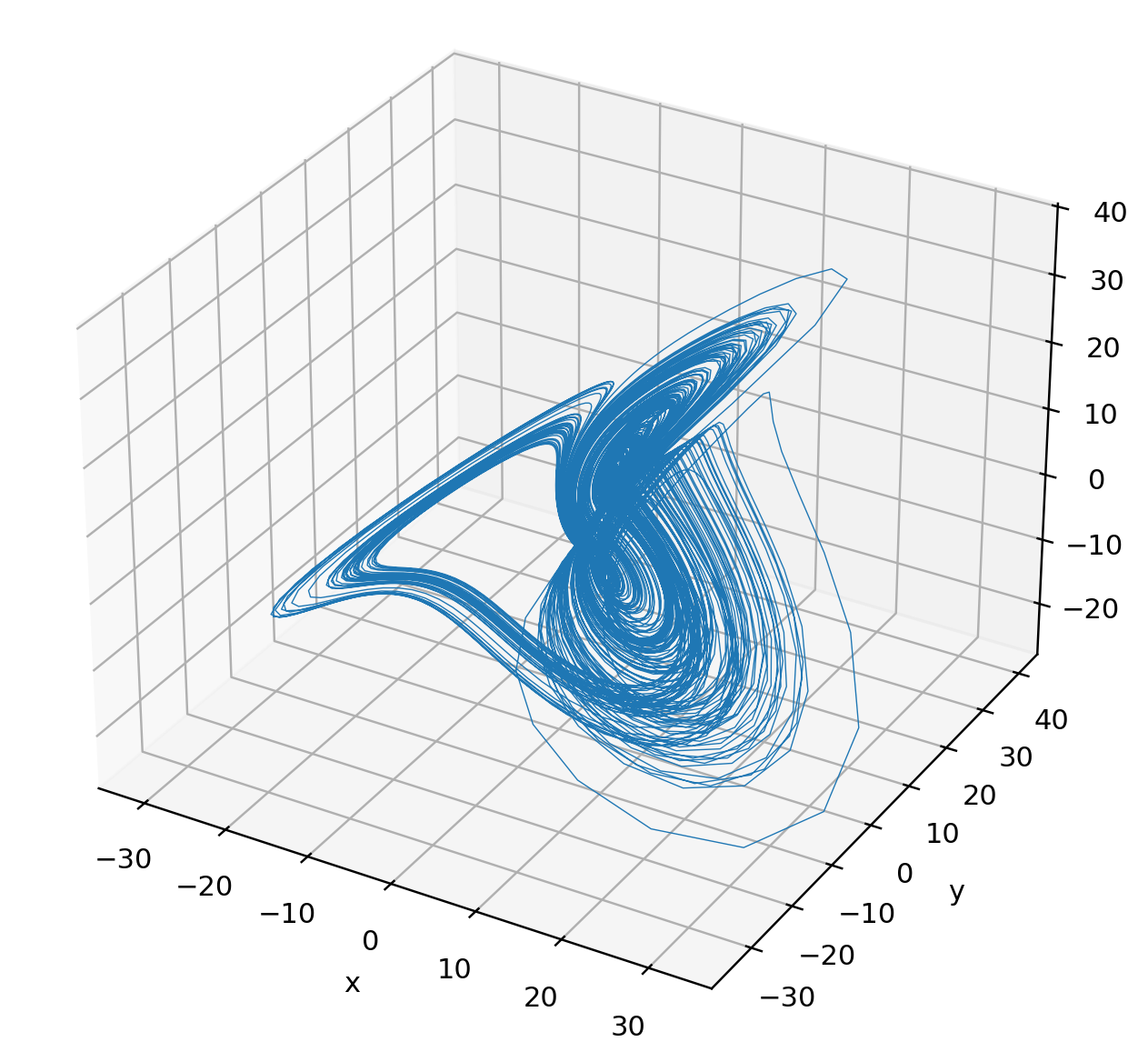}
\caption{Chaotic attractor at $\rho=8$ from $(1,0.5,-0.4)$ after transient removal.}
\label{fig:attractor}
\end{figure}

\section{Fold of Limit Cycles and Intermittent Onset of Chaos}
Continuation of the periodic orbit born at the Hopf point shows that the stable branch does not terminate through period doubling. Instead, its nontrivial Floquet multiplier approaches $+1$. Solving the shooting equations together with the variational dynamics gives a fold of limit cycles at
\begin{equation}
\rhoF\approx4.8150823348,
\label{eq:flc}
\end{equation}
with period $T_{\mathrm{FLC}}\approx1.98187158$. The augmented shooting residual is $7.1\times10^{-15}$. At the fold, two numerically computed multipliers lie within $3.1\times10^{-6}$ of $+1$, while the strongly contracting multiplier is approximately $1.30099\times10^{-7}$. The determinant agrees with the exact identity $\det\Phi(T)=e^{-8T}$.

\begin{figure}[pos=t]
\centering
\includegraphics[width=0.98\columnwidth]{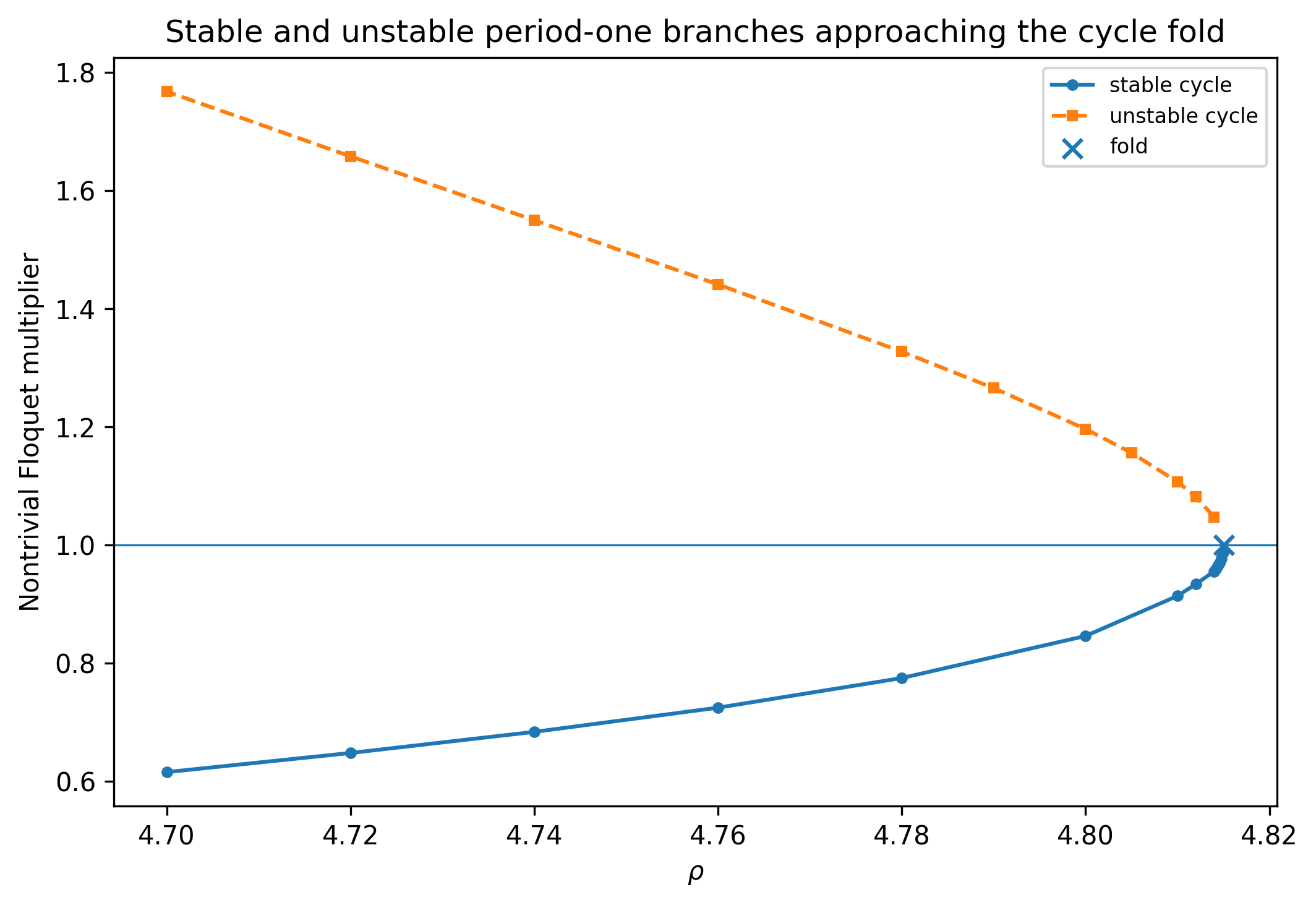}
\caption{Stable and unstable period-one branches approaching the fold of limit cycles through their nontrivial Floquet multipliers.}
\label{fig:floquet}
\end{figure}

The two branches are independently continued below the fold. At $\rho=4.814$, the stable cycle has a nontrivial multiplier $\mu_s\approx0.95568$, whereas the unstable cycle has $\mu_u\approx1.04732$. Figure~\ref{fig:floquet} shows both branches converging to $+1$ at the cyclic saddle-node.

Immediately above \eqref{eq:flc}, trajectories contain long nearly periodic laminar episodes interrupted by irregular bursts. To reduce sensitivity to a small number of parameter samples, we used 25 values logarithmically spaced over
\begin{equation}
4\times10^{-5}\leq\epsilon=\rho-\rhoF\leq5\times10^{-4}.
\end{equation}
After transient removal, laminar episodes were identified from successive maxima satisfying $|x_n-x_{\mathrm{FLC}}|<0.2$, where $x_{\mathrm{FLC}}\approx4.8621$ is the $x$-maximum of the folded periodic orbit. A log--log regression gives
\begin{equation}
\langle\ell\rangle\propto\epsilon^{-0.5510},
\label{eq:scaling}
\end{equation}
with 95\% confidence interval $[-0.6210,-0.4810]$ and $R^2=0.920$. The theoretical type-I value $-1/2$ lies inside this interval \cite{PomeauManneville1980}. A representative laminar-length distribution is also reported to expose the finite-sample variability rather than relying on the fitted exponent alone.

\begin{figure*}[pos=t]
\centering
\subfloat[Scaling based on 25 logarithmically spaced values of $\epsilon$.]{\includegraphics[width=0.43\textwidth]{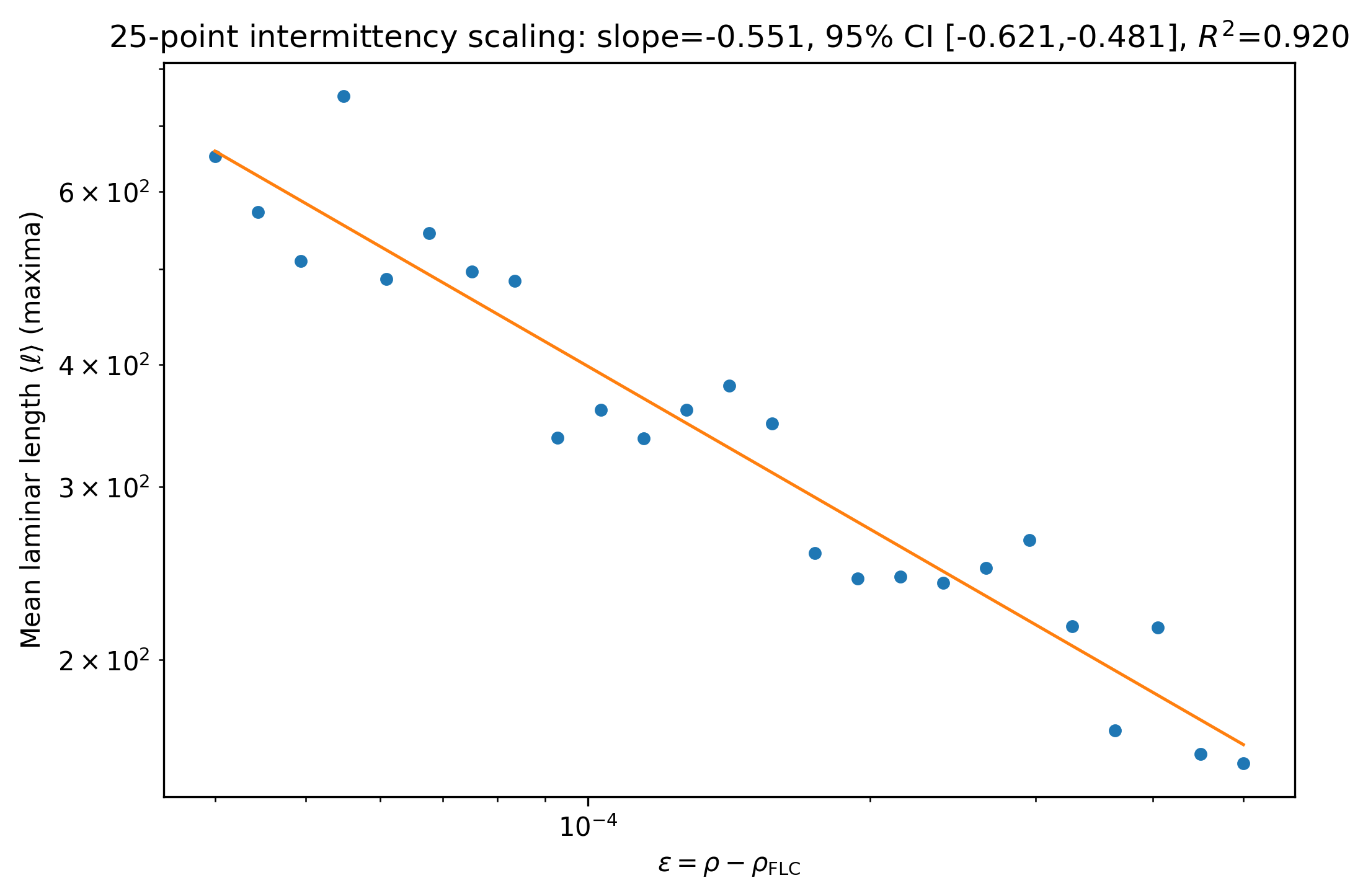}}\hspace{0.04\textwidth}
\subfloat[Representative laminar-length distribution at $\epsilon=2\times10^{-4}$.]{\includegraphics[width=0.43\textwidth]{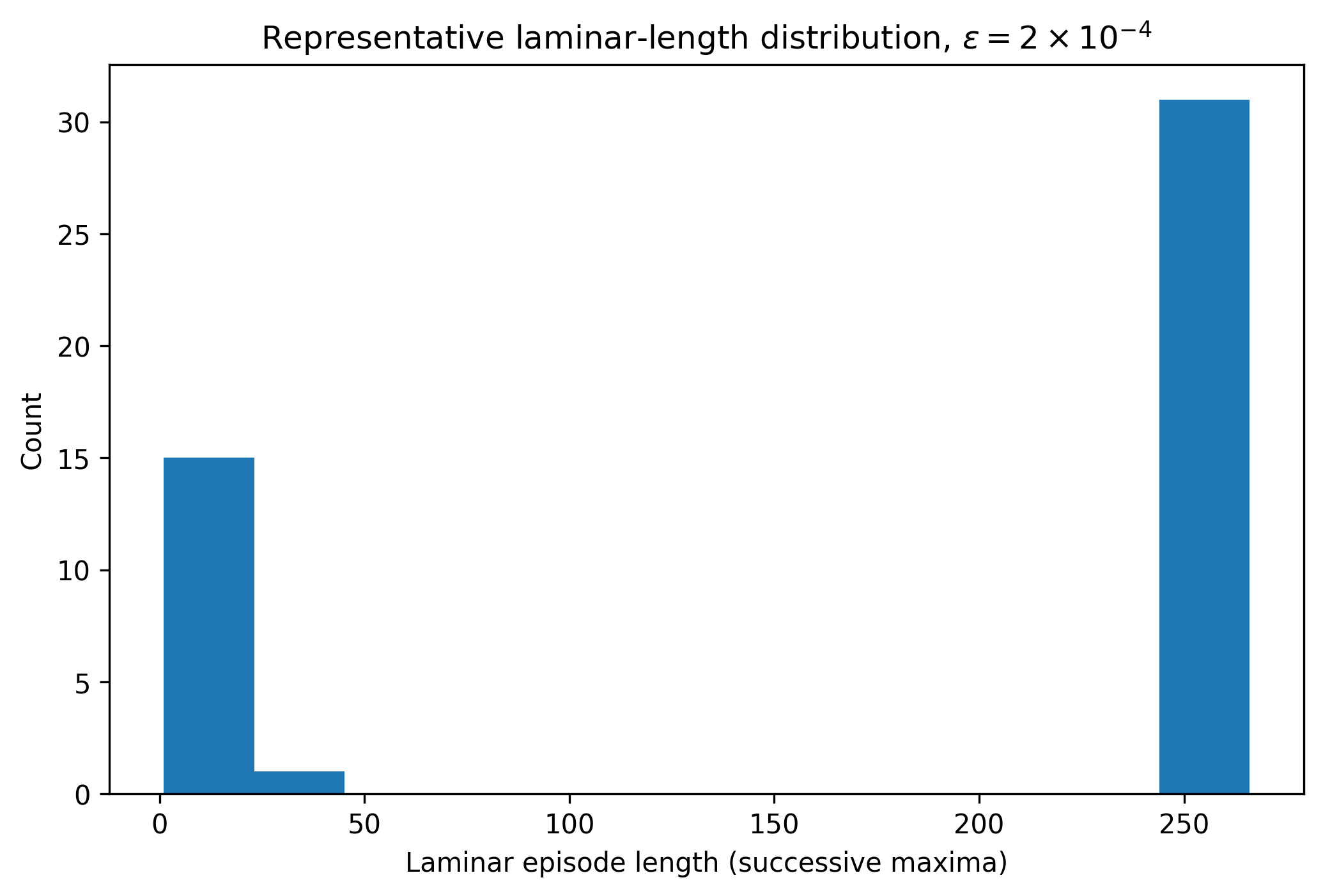}}
\caption{Intermittency analysis near the fold of limit cycles.
	The fitted exponent is $-0.5510$ with 95\% confidence interval
	$[-0.6210,-0.4810]$, consistent with type-I intermittency.}
%\caption{Intermittency analysis near the fold of limit cycles.
%		The fitted exponent is $-0.5510$ with 95\% confidence interval
%		$[-0.6210,-0.4810]$. The result is reported as numerical
%		evidence consistent with type-I intermittency, not as a formal
%		classification theorem.}
\label{fig:intermittency}
\end{figure*}

\section{Chaotic Dynamics and Multistability}
Reference integrations use DOP853 with relative and absolute tolerances $10^{-10}$ and $10^{-12}$ and maximum step $0.01$. Lyapunov exponents are computed from the variational equations with periodic QR reorthogonalization after a long transient.
The numerical results were checked against the exact identities $\sum_i \lambda_i=-8$ and $\det\Phi(T)=e^{-8T}$. Unless stated otherwise, the canonical initial condition is
\begin{equation}
	(x_0,y_0,z_0)=(1,0.5,-0.4).
\end{equation}
At $\rho=6$, the representative spectrum is
\begin{equation}
(\lambda_1,\lambda_2,\lambda_3)\approx(0.846,0,-8.846),
\end{equation}
and at $\rho=8$ it is approximately
\begin{equation}
(\lambda_1,\lambda_2,\lambda_3)\approx(1.04,0,-9.04).
\end{equation}
The sums agree with the exact divergence $-8$, providing a direct numerical consistency check.

\begin{figure}[pos=t]
\centering
\includegraphics[width=0.98\columnwidth]{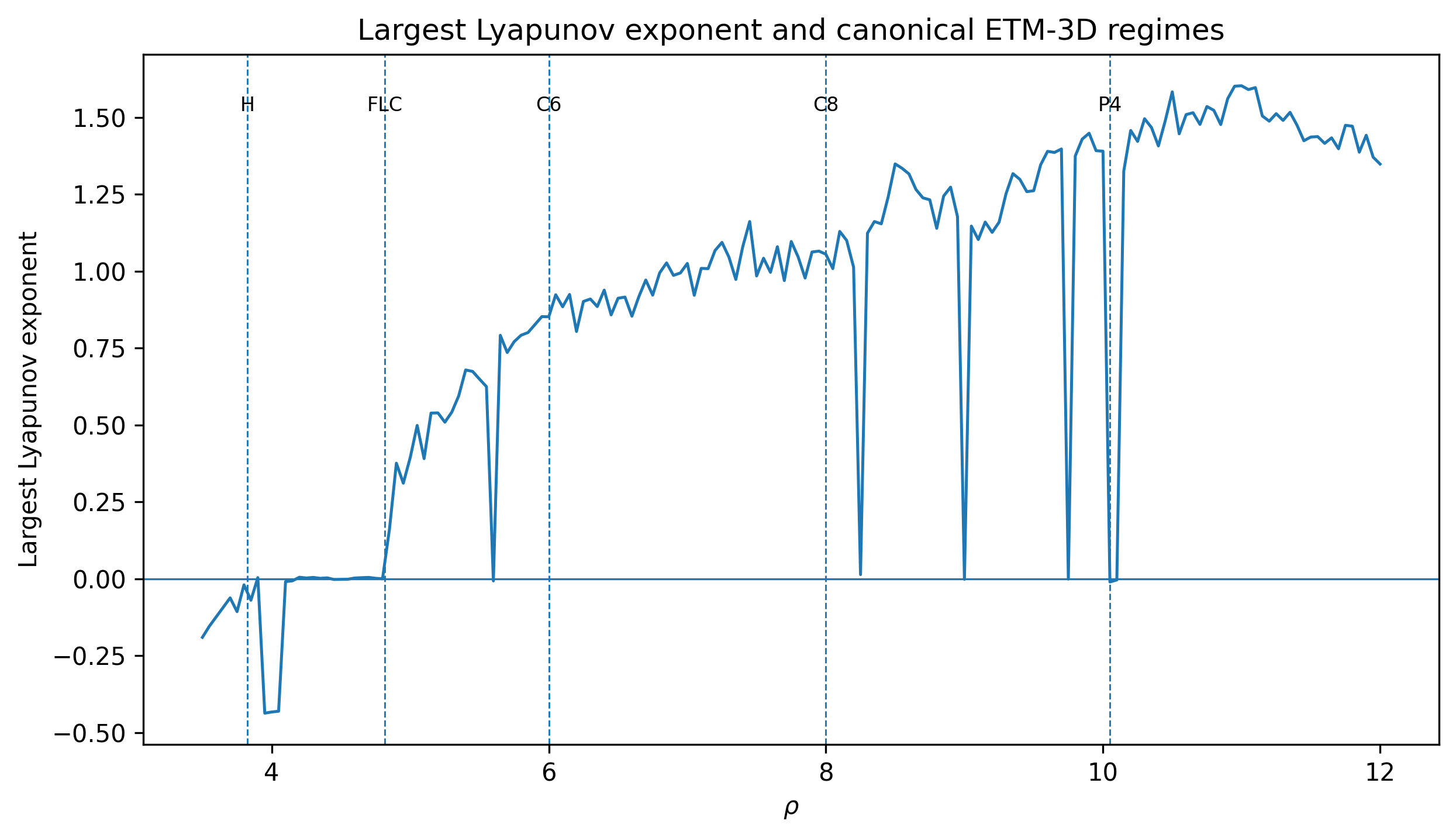}
		\caption{Largest Lyapunov exponent over the studied
			positive-parameter interval. H, FLC, C6, C8, and P4 denote
			the canonical regimes.}
\label{fig:lle}
\end{figure}

At $\rho=8$, the equilibrium
\begin{equation}
E_s\approx(2.68486187,-0.58103484,-1.24165540)
\end{equation}
has eigenvalues
\begin{equation}
-7.41825,\qquad -0.29087\pm4.04347i,
\end{equation}
and is asymptotically stable. It coexists with the chaotic attractor in Fig.~\ref{fig:attractor}; hence the long-term state depends on the initial condition.

\begin{figure*}[pos=t]
\centering
\subfloat[Basin slice at $\rho=8$ with $z(0)=z_s$.]{\includegraphics[width=0.40\textwidth]{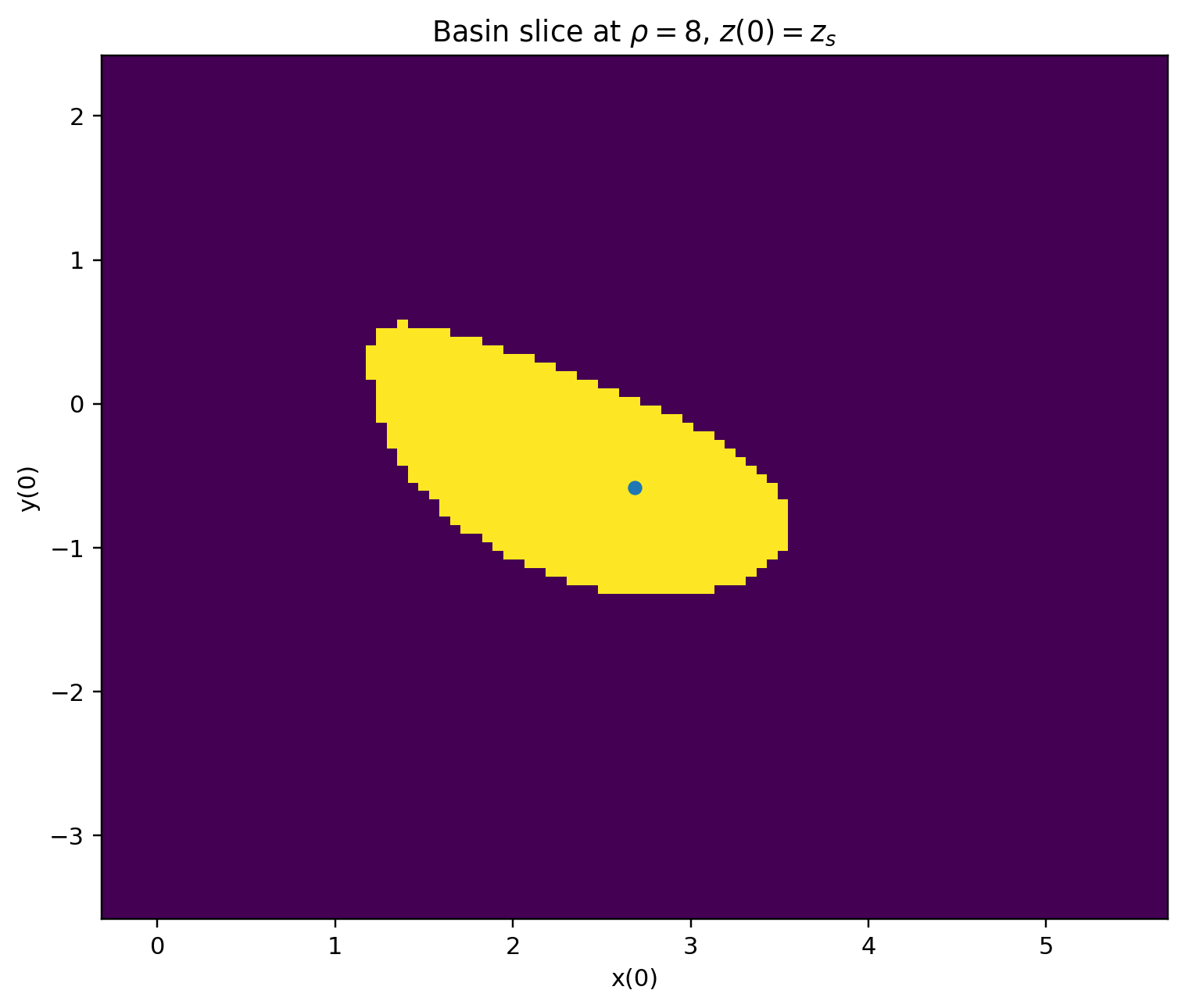}}\hspace{0.05\textwidth}
\subfloat[Return map of successive $x$ maxima on the chaotic attractor.]{\includegraphics[width=0.40\textwidth]{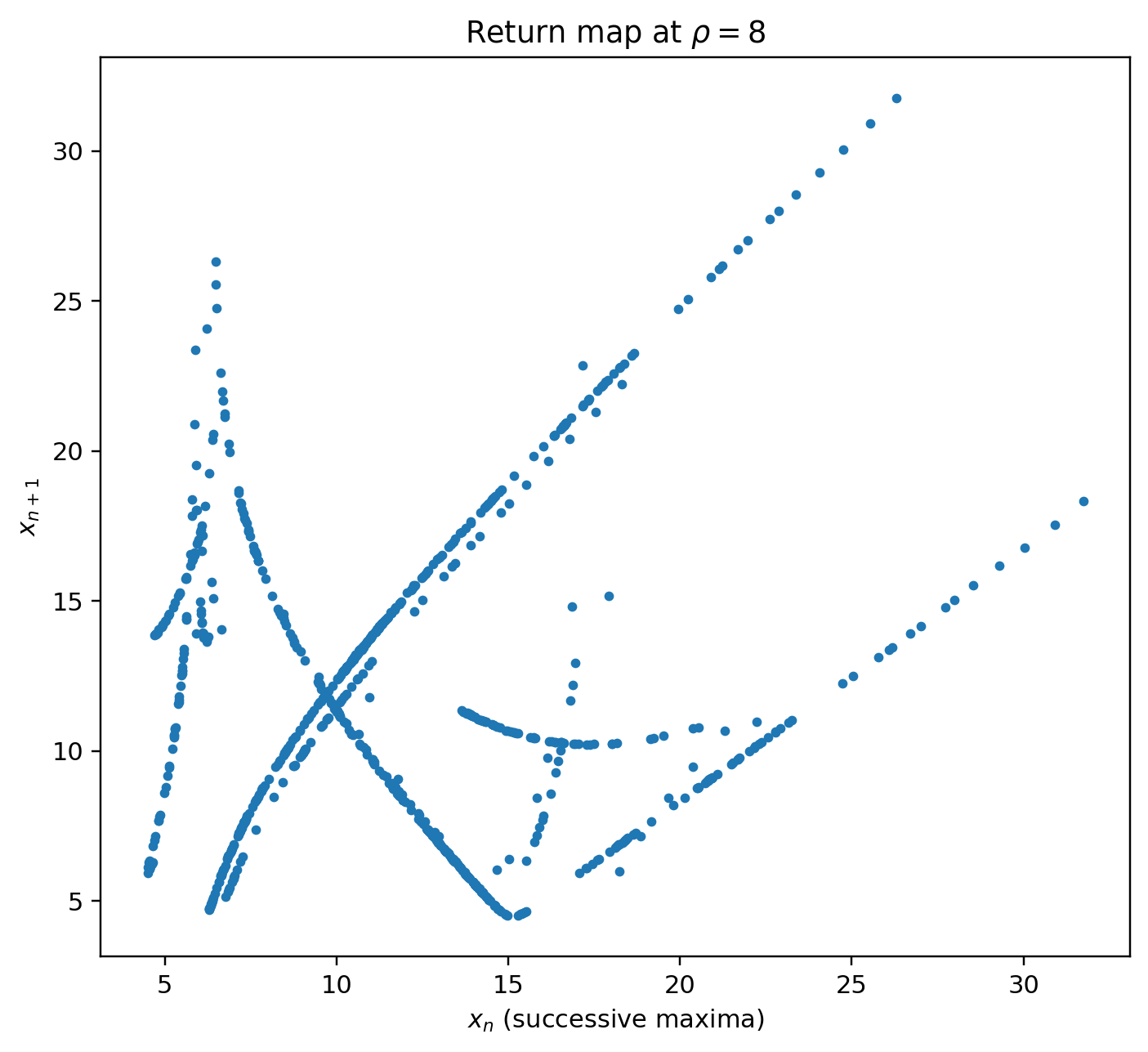}}
\caption{Multistability and return dynamics at $\rho=8$:
		(a) basin slice and (b) return map of successive $x$ maxima.}
\label{fig:basin-return}
\end{figure*}

Periodic windows are present; at $\rho=10.05$, successive $x$ maxima repeat with period four with recurrence error approximately $3.5\times10^{-8}$. ETM-3D is therefore not claimed to possess robust chaos in the strict sense of \cite{Banerjee1998}. Covariant-Lyapunov-vector angle diagnostics at $\rho=8$ also exhibit near tangencies, so no pseudohyperbolicity claim is made \cite{Kuptsov2018}.

\section{Structural Novelty and Equivalence Screening}
A novelty claim based only on the presence of chaos or on the triad $yz$, $xz$, $xy$ would be insufficient. We therefore separate what can be excluded analytically from what remains a broader classification problem.

Consider an invertible affine change of state $x=Pu+c$ and a nonzero time rescaling. Translation changes only the constant and linear terms; the homogeneous quadratic part transforms linearly through $P$. Consequently, the projective zero set of the homogeneous quadratic map is an affine-equivalence invariant up to an invertible linear image.

For ETM-3D,
\begin{equation}
Q_{\mathrm{ETM}}=(yz,-2xz,xy)^T,
\end{equation}
and
\begin{equation}
Q_{\mathrm{ETM}}(x)=0
\quad\Longleftrightarrow\quad
x\in L_x\cup L_y\cup L_z,
\label{eq:qzeros}
\end{equation}
where $L_x,L_y,L_z$ are the three coordinate axes. By contrast, the Lorenz, Chen, and L{\"u} systems have homogeneous quadratic part proportional to $(0,-xz,xy)^T$, whose zero set contains the entire plane $x=0$ in addition to the $x$ axis \cite{Lorenz1963,ChenUeta1999,Lu2002}. Since an invertible linear map cannot transform a two-dimensional component into a union of one-dimensional components, ETM-3D is not affinely equivalent to these Lorenz-type systems.

The Qi, Rabinovich, and Newton--Leipnik systems contain three nonzero cross products \cite{Qi2005,Rabinovich1978,LeipnikNewton1981}; their homogeneous quadratic zero set has the same three-axis structure. In this case any linear equivalence of the quadratic cores must map the three axes onto the three axes, and therefore the transformation matrix is monomial (a coordinate permutation followed by diagonal scaling). For $\rho>0$, the ETM-3D linear matrix
\begin{equation}
A_{\mathrm{ETM}}(\rho)=
\begin{bmatrix}
-2&-\rho&0\\
1&-1&\rho\\
0&\rho&-5
\end{bmatrix}
\end{equation}
has seven nonzero entries, whereas the canonical linear matrices of the Qi, Rabinovich, and Newton--Leipnik forms have five. Monomial similarity and time rescaling preserve this zero pattern up to permutation. Hence there is no origin-preserving scaling/permutation/time-rescaling equivalence between ETM-3D with $\rho>0$ and those canonical forms.

\begin{table*}[pos=t]
\caption{Structural Comparison With Representative Quadratic Chaotic Families}
\label{tab:equiv}
\centering
\footnotesize
\resizebox{\textwidth}{!}{%
\begin{tabular}{@{}llll@{}}
\toprule
Family & Homogeneous quadratic structure & Screening invariant & Conclusion of present test\\
\midrule
Lorenz / Chen / L{\"u} & two cross terms & quadratic zero set contains a plane & arbitrary affine equivalence excluded\\
Qi & three cross terms & three-axis zero set; monomial map required & origin-preserving scaling equivalence excluded\\
Rabinovich & three cross terms & three-axis zero set; monomial map required & origin-preserving scaling equivalence excluded\\
Newton--Leipnik & three cross terms & three-axis zero set; monomial map required & origin-preserving scaling equivalence excluded\\
Sprott A--S & simple quadratic catalogue & canonical-form literature screen & no complete affine classification claimed\\
\bottomrule
\end{tabular}%
}
\end{table*}

Sprott's catalogue contains 19 distinct simple quadratic chaotic flows \cite{Sprott1994}. We include it as a systematic canonical-form literature screen, but do not claim that Table~\ref{tab:equiv} constitutes a complete algebraic classification against every possible translated affine representation of every Sprott flow. Likewise, for the full-cross triad families, an unrestricted affine conjugacy that maps the ETM origin to a nonzero equilibrium would alter the lower-order terms and is not ruled out by the zero-pattern argument alone. Accordingly, the originality statement is deliberately limited to the exact architecture and the equivalence tests established above rather than an absolute worldwide non-equivalence claim.

\section{Canonical Parameter Regimes}

Because the dynamics of ETM-3D vary markedly with $\rho$,
Table~\ref{tab:bench} defines representative parameter values for
distinct regimes within the same one-parameter system. ETM-H and
ETM-F correspond to the supercritical Hopf point and the fold of
limit cycles, respectively. ETM-I represents the intermittent
post-fold regime, while ETM-C6 and ETM-C8 correspond to developed
chaos, with ETM-C8 additionally exhibiting chaos--equilibrium
multistability. ETM-P4 denotes the period-four window near
$\rho=10.05$.

\begin{table}[!h]
	\caption{Canonical ETM-3D parameter regimes}
	\label{tab:bench}
	\centering
	\footnotesize
	\begin{tabular}{@{}lll@{}}
		\toprule
		ID & $\rho$ & Dynamical regime \\
		\midrule
		ETM-H  & 3.82517466 & Hopf point \\
		ETM-F  & 4.8150823348 & Fold of limit cycles \\
		ETM-I  & 4.81520 & Intermittent onset of chaos \\
		ETM-C6 & 6.0 & Developed chaos \\
		ETM-C8 & 8.0 & Chaos--equilibrium multistability \\
		ETM-P4 & 10.05 & Period-four window \\
		\bottomrule
	\end{tabular}
\end{table}

\section{Discussion}

ETM-3D combines a simple one-parameter architecture with a rich
organization of nonlinear behavior. Its quadratic part preserves
Euclidean quadratic energy, while the full vector field has constant
negative divergence. This separation permits a global boundedness
result through an explicit positively invariant absorbing set,
independent of numerical attractor reconstruction. 
The positive-parameter bifurcation structure provides a consistent
route from equilibria to periodic and chaotic motion. Generic
equilibria reduce to a scalar quartic, the singular case at
$\rho=\sqrt{5}$ is treated separately, and the transcritical and Hopf
bifurcations satisfy explicit nondegeneracy conditions. The stable
periodic orbit created at the supercritical Hopf point terminates at
a fold of limit cycles, where it collides with an unstable branch. 
Beyond the fold, the laminar-length scaling is consistent with
type-I intermittency, although this classification remains numerical.
At $\rho=8$, a stable equilibrium coexists with the chaotic attractor,
demonstrating multistability, while the period-four window near
$\rho=10.05$ shows that chaos is interrupted. Accordingly, ETM-3D is
not claimed to exhibit robust chaos, and the observed CLV
near-tangencies do not support pseudohyperbolicity. 
The originality of ETM-3D is associated with its complete
one-parameter linear--quadratic architecture rather than with the
quadratic triad alone. The structural tests distinguish it from the
classical families considered in Section~6 within the stated
equivalence classes, without claiming an exhaustive affine
classification. With no direct physical derivation, ETM-3D is
intended as a compact mathematical benchmark for bifurcation,
multistability, intermittency, and chaotic-dynamics studies.
\section{Conclusion}
A one-parameter three-dimensional quadratic flow has been introduced
and analyzed on the benchmark domain $\rho\geq0$. The system combines
an energy-preserving quadratic interaction with constant
phase-space contraction and admits an explicit positively invariant,
globally absorbing set. Its equilibrium structure can be reduced
analytically to a scalar quartic apart from the singular case at
$\rho=\sqrt{5}$, and the positive parameter branch contains two
saddle-node bifurcations, a generic transcritical bifurcation, and a
supercritical Hopf bifurcation. 
Continuation of the periodic solution reveals a fold of stable and
unstable limit cycles, followed by a post-fold regime whose
laminar-length scaling is consistent with type-I intermittency.
Further increases in $\rho$ produce developed chaos, coexistence of a
chaotic attractor with a stable equilibrium, and periodic windows.
The resulting dynamics are therefore multistable and
parameter-dependent rather than uniformly chaotic.
 Together with the structural equivalence screening, these results
identify ETM-3D as a compact and reproducible quadratic system in
which local bifurcations, periodic-orbit transitions, intermittency,
multistability, and chaos can be studied within a single
one-parameter family. Future work may extend the bifurcation analysis
to the negative parameter branch, investigate more general affine
equivalence classes, and evaluate the canonical regimes in
forecasting, identification, and nonlinear-control applications.

\section*{Declaration of AI-assisted technologies in the manuscript preparation process}
During the preparation of this work, the authors used OpenAI ChatGPT solely to assist with language editing and improvement of the written presentation of the manuscript. The authors reviewed and edited the text and take full responsibility for the content of the article.


\begin{thebibliography}{20}
\bibitem{Lorenz1963}
E.~N. Lorenz, ``Deterministic nonperiodic flow,'' \emph{J. Atmos. Sci.}, vol.~20, no.~2, pp.~130--141, Mar. 1963, doi: 10.1175/1520-0469(1963)020$<$0130:DNF$>$2.0.CO;2.

\bibitem{ChenUeta1999}
G. Chen and T. Ueta, ``Yet another chaotic attractor,'' \emph{Int. J. Bifurcation Chaos}, vol.~9, no.~7, pp.~1465--1466, 1999, doi: 10.1142/S0218127499001024.

\bibitem{Lu2002}
J. L{\"u}, G. Chen, D. Cheng, and S. Celikovsky, ``Bridge the gap between the Lorenz system and the Chen system,'' \emph{Int. J. Bifurcation Chaos}, vol.~12, no.~12, pp.~2917--2926, 2002, doi: 10.1142/S021812740200631X.

\bibitem{LeipnikNewton1981}
R.~B. Leipnik and T.~A. Newton, ``Double strange attractors in rigid body motion with linear feedback control,'' \emph{Phys. Lett. A}, vol.~86, no.~2, pp.~63--67, Nov. 1981, doi: 10.1016/0375-9601(81)90165-1.

\bibitem{Rabinovich1978}
M.~I. Rabinovich, ``Stochastic self-oscillations and turbulence,'' \emph{Sov. Phys. Usp.}, vol.~21, pp.~443--469, 1978, doi: 10.1070/PU1978v021n05ABEH005555.

\bibitem{Sprott1994}
J.~C. Sprott, ``Some simple chaotic flows,'' \emph{Phys. Rev. E}, vol.~50, no.~2, pp.~R647--R650, Aug. 1994, doi: 10.1103/PhysRevE.50.R647.

\bibitem{Qi2005}
G. Qi, G. Chen, S. Du, Z. Chen, and Z. Yuan, ``Analysis of a new chaotic system,'' \emph{Physica A}, vol.~352, nos.~2--4, pp.~295--308, Jul. 2005, doi: 10.1016/j.physa.2004.12.040.

\bibitem{PomeauManneville1980}
Y. Pomeau and P. Manneville, ``Intermittent transition to turbulence in dissipative dynamical systems,'' \emph{Commun. Math. Phys.}, vol.~74, no.~2, pp.~189--197, 1980, doi: 10.1007/BF01197757.

\bibitem{Banerjee1998}
S. Banerjee, J.~A. Yorke, and C. Grebogi, ``Robust chaos,'' \emph{Phys. Rev. Lett.}, vol.~80, no.~14, pp.~3049--3052, Apr. 1998, doi: 10.1103/PhysRevLett.80.3049.

\bibitem{Kuptsov2018}
P.~V. Kuptsov and S.~P. Kuznetsov, ``Lyapunov analysis of strange pseudohyperbolic attractors: Angles between tangent subspaces, local volume expansion and contraction,'' \emph{Regular Chaotic Dyn.}, vol.~23, nos.~7--8, pp.~908--932, 2018, doi: 10.1134/S1560354718070079.

\bibitem{Kuznetsov2004}
Y.~A. Kuznetsov, \emph{Elements of Applied Bifurcation Theory}, 3rd~ed. New York, NY, USA: Springer, 2004, doi: 10.1007/978-1-4757-3978-7.

\bibitem{Gilpin2021}
W. Gilpin, ``Chaos as an interpretable benchmark for forecasting and data-driven modelling,'' in \emph{Proc. Neural Information Processing Systems Track on Datasets and Benchmarks}, vol.~1, 2021.

\bibitem{Gkimisis2026}
L. Gkimisis, I.~P. Duff, P. Goyal, and P. Benner, ``On the representation of energy-preserving quadratic operators with application to Operator Inference,'' \emph{Appl. Math. Lett.}, vol.~173, Art. no.~109761, Feb. 2026, doi: 10.1016/j.aml.2025.109761.

\bibitem{Amirabadi2021}
R. Khalili Amirabadi, O. S. Fard, and A. Mansoori,
``A novel fuzzy sliding mode control approach for chaotic systems,''
\emph{Iranian J. Fuzzy Syst.}, vol.~18, no.~6, pp.~133--150, 2021,
doi: 10.22111/ijfs.2021.6338.

\bibitem{Amirabadi2026}
R. Khalili Amirabadi, O. S. Fard, A. Mansoori, and M. Jalaeian-Farimani,
``Improved control of uncertain chaotic systems using hybrid fuzzy sliding mode techniques: Chattering elimination and robust performance,''
\emph{Fuzzy Sets Syst.}, vol.~542, Art. no.~109944, 2026,
doi: 10.1016/j.fss.2026.109944.

\end{thebibliography}
\end{document}